\documentclass[apj]{emulateapj}
\usepackage{natbib}
\usepackage{rotating}

\newcommand{\myemail}{maravena@nrao.edu}

\shorttitle{Cold molecular gas in massive disk galaxies at $z=1.5$}
\shortauthors{Aravena et al.}

\begin{document}


\title{Cold molecular gas in massive, star-forming disk galaxies at $z=1.5$}


\author{M. Aravena\altaffilmark{1}, C. Carilli\altaffilmark{1},  E. Daddi\altaffilmark{2}, J. Wagg\altaffilmark{3}, F. Walter\altaffilmark{4}, D. Riechers\altaffilmark{5,6}, H. Dannerbauer\altaffilmark{2}, G. E. Morrison\altaffilmark{7,8}, D. Stern\altaffilmark{9}, M. Krips\altaffilmark{10}}

\altaffiltext{$^\ast$}{The National Radio Astronomy Observatory is a facility of the National Science Foundation (NSF), operated under cooperative agreement by Associated Universities Inc.}
\altaffiltext{1}{National Radio Astronomy Observatory, P.O. Box O, Socorro, NM 87801, USA. \myemail}
\altaffiltext{3}{CEA Saclay, Laboratoire AIM, Irfu/SAp, Orme des Merisiers, F-91191 Gif-sur-Yvette Cedex, France}
\altaffiltext{3}{European Southern Observatory, Alonso de C\'ordoba 3107, Vitacura Santiago, Chile}
\altaffiltext{4}{Max-Planck-Institut f\"ur Astronomie, K\"onigstuhl 17, D-69117 Heidelberg, Germany}
\altaffiltext{5}{California Institute for Technology, Pasadena, CA 91109}
\altaffiltext{6}{Hubble Fellow}
\altaffiltext{7}{Institute for Astronomy, University of Hawaii, Honolulu, HI 96822}
\altaffiltext{8}{Canada-France-Hawaii Telescope, Kamuela, HI 96743}
\altaffiltext{9}{Jet Propulsion Laboratory, California Institute for Technology, Pasadena, CA 91109}
\altaffiltext{10}{Institut de Radio Astronomie Millim\'etrique (IRAM), St. Martin d'Heres, France}


\begin{abstract}
We report the detection of the CO $J=1-0$ emission line in three near-infrared selected star-forming galaxies at $z\sim1.5$ with the Very Large Array (VLA) and the Green Bank telescope (GBT). These observations directly trace the bulk of molecular gas in these galaxies. We find H$_2$ gas masses of $8.3\pm1.9\times10^{10}\ M_\sun$, $5.6\pm1.4\times 10^{10}\ M_\sun$ and $1.23\pm0.34\times10^{11}\ M_\sun$ for BzK-4171, BzK-21000 and BzK-16000, respectively, assuming a conversion $\alpha_\mathrm{CO}=3.6\ M_\sun$ (K km s$^{-1}$ pc$^{2}$)$^{-1}$. We combined our observations with previous CO $2-1$ detections of these galaxies to study the properties of their molecular gas.  
We find brightness temperature ratios between the CO $2-1$ and CO $1-0$ emission lines of $0.80_{-0.22}^{+0.35}$, $1.22_{-0.36}^{+0.61}$ and $0.41_{-0.13}^{+0.23}$ for BzK-4171, BzK-21000 and BzK-16000, respectively. At the depth of our observations it is not possible to discern between thermodynamic equilibrium or sub-thermal excitation of the molecular gas at $J=2$. However, the low temperature ratio found for BzK-16000 suggests sub-thermal excitation of CO already at $J=2$. For BzK-21000, a Large Velocity Gradient model of its CO emission confirms previous results of the low-excitation of the molecular gas at $J=3$. From a stacked map of the  CO $1-0$ images, we measure a CO $2-1$ to CO $1-0$ brightness temperature ratio of $0.92_{-0.19}^{+0.28}$. This suggests that, on average, the gas in these galaxies is thermalized up to $J=2$, has star-formation efficiencies of $\sim100$ $L_\sun$ (K km s$^{-1}$ pc$^2$)$^{-1}$ and gas consumption timescales of $\sim0.4$ Gyr, unlike SMGs and QSOs at high redshifts. 

\end{abstract}


\keywords{galaxies: evolution --- galaxies: formation --- cosmology: observations --- galaxies: starburst --- galaxies: high-redshift}




\section{Introduction}

\begin{table*}
\centering
\caption{Summary of the VLA observations\label{table:1}}
\begin{tabular}{lcccccrcc}
\hline
 Source & R.A.$^a$ & Dec.$^a$ & Frequency$^b$ & Bandwidth & Beam & Pos. Ang. & Cov. Fraction$^c$ & Rms$^d$ \\
        & (J2000) & (J2000) & (GHz) & (MHz) &   &  & (\%) &($\mu$Jy/beam) \\
\hline\hline
BzK-4171 & 12\ 36\ 26.516 & $+$62\ 08\ 35.35   & 46.760 & 100  & $1.69\arcsec\times1.56\arcsec$ & $+09.8\degr$  &   96       &  44 \\
         &                &                    & 43.340 & 100 & $2.02\arcsec\times1.67\arcsec$ & $-53.0\degr$   &          & 62 \\
BzK-21000 & 12\ 37\ 20.597 & $+$62\ 22\ 34.60   & 45.710 & 100  & $1.79\arcsec\times1.61\arcsec$ & $+18.2\degr$ &   98       & 42 \\
         &                &                    & 43.340 & 100 & $2.03\arcsec\times1.67\arcsec$ & $-54.2\degr$   &          &72 \\
BzK-16000 & 12\ 36\ 30.120 & $+$62\ 14\ 28.00   & 45.653 & $13\times3.125$ & $1.96\arcsec\times1.60\arcsec$  & $+48.9\degr$ &  76    & 145\\
         &                &                    & 45.635$^e$ & 100  & $1.90\arcsec\times1.62\arcsec$ & $+44.3\degr$ &        & 181 \\
\hline
\end{tabular}
\begin{flushleft}
\noindent $^a$ VLA 1.4 GHz position from \citet{Morrison2010}. 
\noindent $^b$ Central frequency between the two 50 MHz IFs. 
\noindent $^c$ Fraction of the CO line covered by the VLA observations. $^d$ Image noise level measured around the source position. $^e$ Combined continuum emission from two 50 MHz bandwidth channels at 45.585 GHz and 45.685 GHz.\end{flushleft}
\end{table*}

\begin{figure*}
\centering
 \includegraphics[scale=0.38]{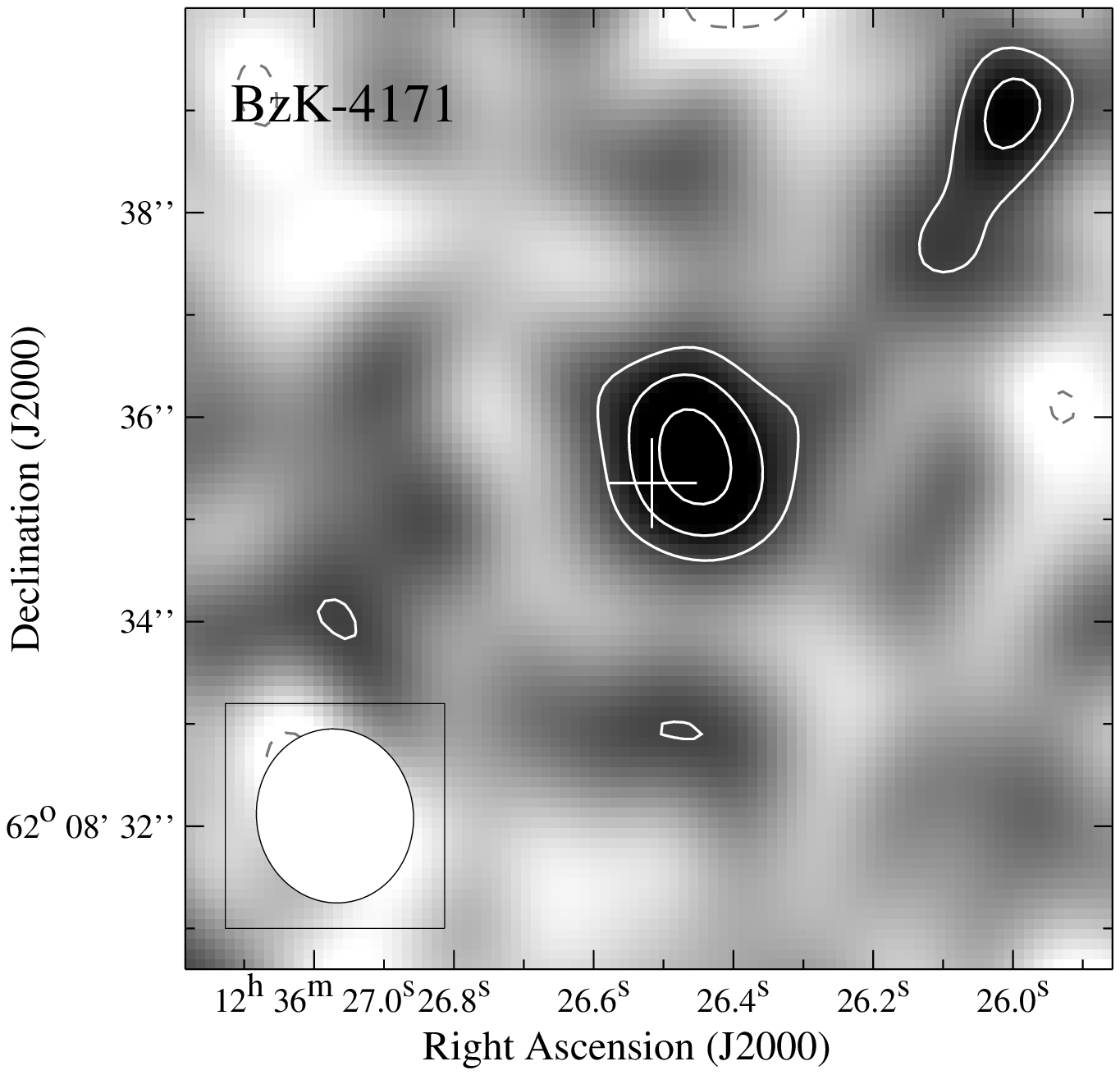}
\includegraphics[scale=0.38]{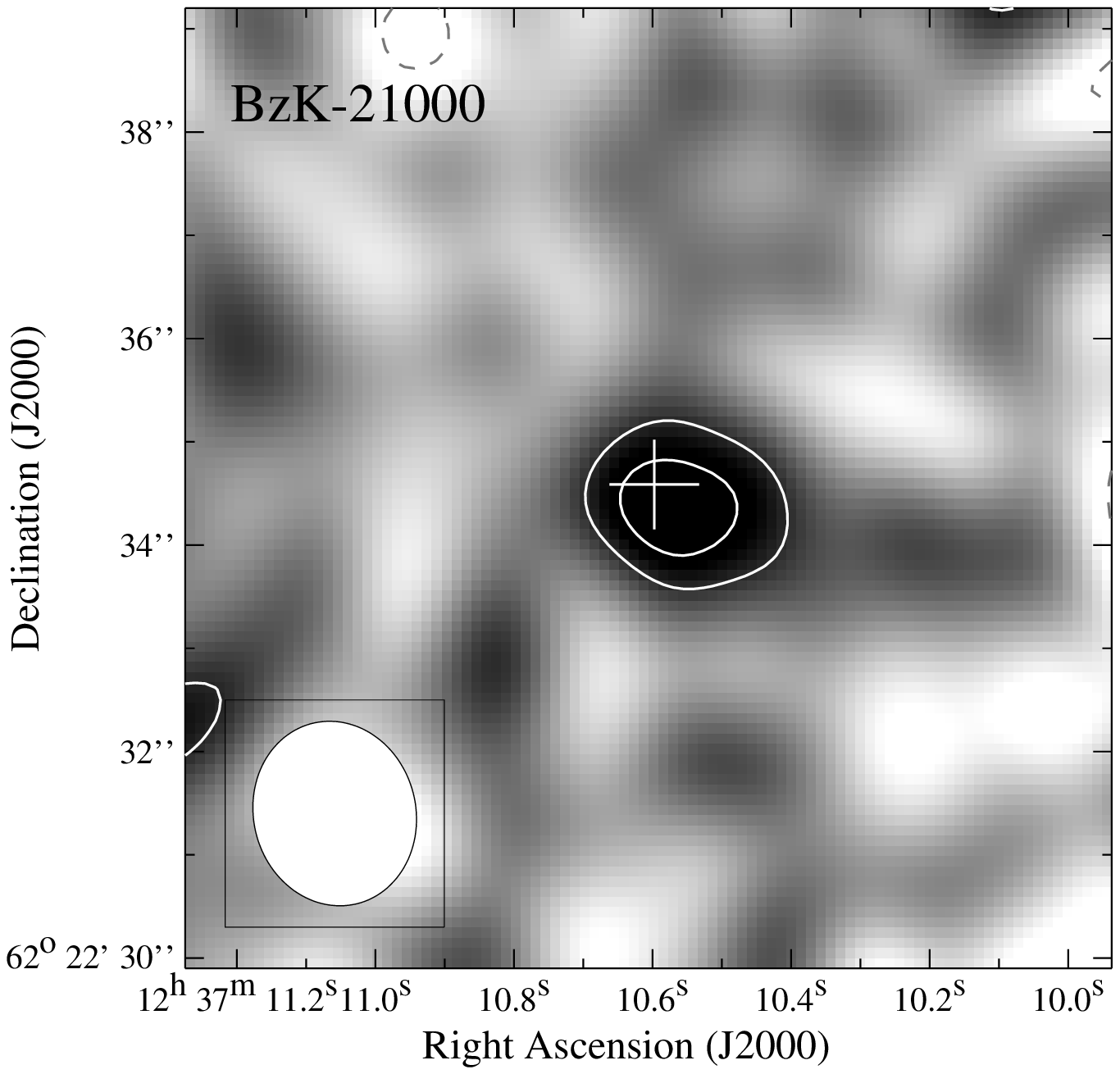}
 \includegraphics[scale=0.38]{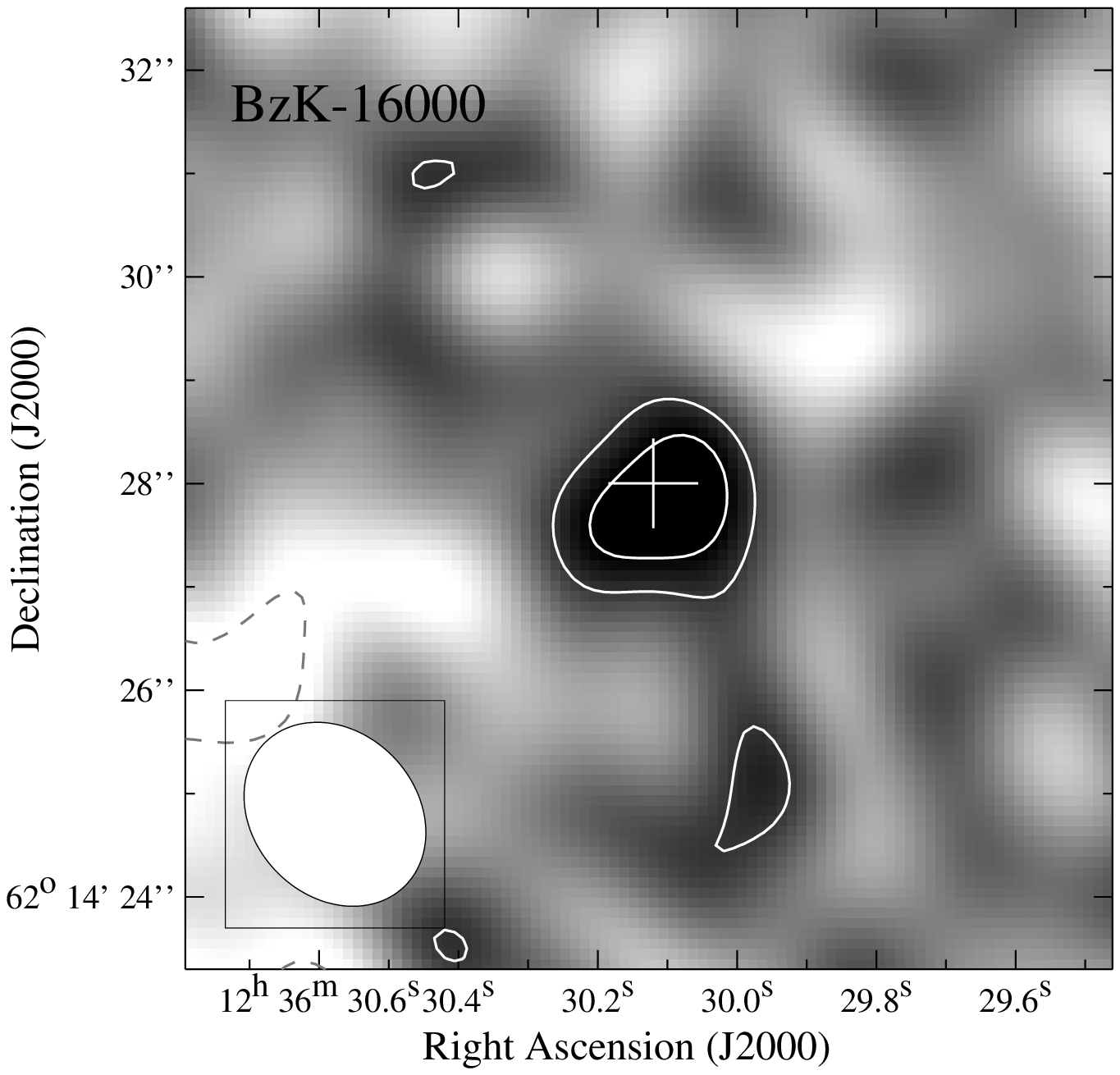}
\caption{VLA images of the CO $1-0$ emission line from our sources. Contour levels are: -2, 2, 3 and 4$\sigma$. The crosses indicate the VLA 1.4 GHz position for our sources \citep{Morrison2010}. \label{fig:bzk_im}}
\end{figure*}

A remarkable step in our understanding of galaxy formation has come from the determination of the unobscured history of the star formation rate (SFR) density over a wide range of redshift. The SFR history of the Universe has a peak from $z=3$ to 1, and presents a steady decline at $z<1$ \citep[e.g.][]{Lilly1996, Madau1996, Steidel1999}. Apparently, a significant contribution to this peak is provided by the vigorous bursts of star formation triggered by galaxy mergers and interactions. In these cases, the star formation is very efficient, showing high SFRs and high gas surface densities on scales of several kpc, similar to what is observed in distant submillimeter galaxies \citep[SMGs;][]{Tacconi2006,Tacconi2008} and quasars \citep[QSOs;][]{Walter2004, Walter2009, Riechers2008, Riechers2009}. However, an important fraction of the SFR density appears to come from normal star-forming disk galaxies \citep{Bell2005, Elbaz2007, Genzel2008}. Here, continuous flows of cold gas from the intergalactic medium may provide the necessary fuel for star formation \citep{Keres2005, Dekel2009}. 

Studies of the properties of the molecular gas (e.g., CO) in galaxies during the main epoch of galaxy formation ($z=1-3$) have mostly focused on SMGs and QSOs due to the limited capabilities of millimeter and radio telescopes and instruments \citep[e.g.,][]{Solomon2005, Weiss2005b, Riechers2006, Carilli2007, Weiss2007, Aravena2008, Coppin2008, Carilli2010}. These IR-luminous objects were found to have large reservoirs of molecular gas ($M(\mathrm{H}_2)\sim10^{10-11} M_\sun$) that sustain SFRs of $\sim500-1000\ M_\sun$ yr$^{-1}$ for $<100$ Myr. Similar to what is observed in local ultra-luminous infrared galaxies (ULIRGs), the molecular gas is in local thermodynamic equilibrium (LTE) up to high-$J$ CO transitions \citep[$J>3$;][]{Riechers2006, Weiss2007}. This is expected for warm, dense H$_2$ molecular gas. 

Recently, large amounts of molecular gas, similar to those observed in bright high-redshift SMGs and QSOs, were found in six relatively quiescent disk galaxies at $z\sim1.5$ \citep{Daddi2008, Daddi2009}, 14 similar galaxies at $z\sim1.2$ and $z\sim2$ \citep{Tacconi2010} and three disk galaxies at $z=0.5$ \citep[][; Salmi et al. in preparation]{Daddi2010b}. A detailed analysis of the properties of these near-IR selected galaxies revealed SFRs $\sim50-200\ M_\sun$ yr$^{-1}$, with stellar masses of $\sim10^{10-11}\ M_\sun$. These galaxies apparently have high molecular gas fractions, low star-formation efficiencies (SFEs) \citep{Daddi2008, Daddi2009,Tacconi2010} and CO luminosity to gas mass conversion factors similar to that found in the Milky Way galaxy disk rather than that of starbursting ULIRGs and SMGs \citep{Daddi2009}. This means that these objects have yet to convert a large fraction of their gas into stars, indicating galaxies right in the process of stellar build-up. The first CO $1-0$ measurements in two BzK galaxies were previously reported by \citet{Dannerbauer2009}, but only led to a stacked detection at $\sim3\sigma$. Those observations, combined with the first detection of the CO $3-2$ line in one of these galaxies (BzK-21000), allowed \citet{Dannerbauer2009} to perform an analysis of the excitation properties of the molecular gas, finding low excitation conditions, similar to what is seen in local disk galaxies. This suggests an important difference to what is found in more luminous objects at high-redshift, where the CO emission is in LTE up to high$-J$ CO transitions. 

In this paper, we report individual detections of the CO $J=1-0$ emission line in a sample of three near-IR selected star-forming galaxies at $z\sim1.5$ that have previously been studied in detail by \citet{Daddi2009}: BzK-4171, BzK-21000 and BzK-16000. In section \ref{section:obs} we describe our CO $1-0$ observations. In section \ref{section:res}, we present our results, give fluxes and luminosities. In section \ref{section:diss}, we summarize our results and discuss the average properties of the molecular gas in our galaxies. We assume a concordance cosmology with $H_0=71$ km s$^{-1}$ Mpc$^{-1}$, $\Omega_\lambda=0.73$ and $\Omega_m=0.27$.

\section{Observations}
\label{section:obs}

\begin{table*}
\centering
\caption{Properties of the molecular gas.\label{table:properties}}
 \begin{tabular}{lccccccc}
\hline
Source & $S_{43 \mathrm{GHz}}^a$ & $S_{\mathrm{CO}\ 1-0}^b$ & $I_{\mathrm{CO}\ 1-0}^c$ & $L_\mathrm{CO}'\ ^d$ & $M(\mathrm{H}_2)^e$ &  $t_\mathrm{gas}^f$ & SFE$^g$\\
       &  ($\mu$Jy)              & ($\mu$Jy)                & (Jy km s$^{-1}$)       & ($\times10^{10}$ K km s$^{-1}$ pc$^2$) & ($\times10^{10} M_\sun$) & ($\times10^9$ yr) & ($L_\sun$ (K km s$^{-1}$ pc$^2$)$^{-1}$)\\
\hline\hline
BzK-4171 & $<125$ & $305\pm70$ & $0.20\pm0.05$ & $2.31\pm0.53$ & $8.3\pm1.9$  & 0.80 & 43 \\
BzK-21000& $<144$ & $180\pm50$ & $0.13\pm0.03$ & $1.55\pm0.40$ & $5.6\pm1.4$  & 0.25 & 142\\
BzK-16000& $<360$ & $805\pm220$ & $0.28\pm0.08$ & $3.42\pm0.94$ & $12.3\pm3.4$ & 0.81 & 44 \\
\hline
\end{tabular}\\
\begin{flushleft}
\noindent $^a$ Upper limit (2$\sigma$) to the continuum emission at 43.34 GHz for BzK-4171 and BzK-21000 and at 45.6 GHz for BzK-16000. $^b$ Flux density integrated over an area that matches the CO $2-1$ measurements. $^c$ Velocity integrated intensity, $I=\int S dv$, corrected for emission that falls outside the range covered by the VLA band. $^d$ CO luminosity. $^e$ H$_2$ mass computed using a CO luminosity to gas mass conversion factor of 3.6 (K km s$^{-1}$ pc$^{2}$)$^{-1}$ \citep{Daddi2009}. $^f$ Gas depletion lifetimes and $^g$ star formation efficiencies derived using the SFR and the far-IR luminosities from \citet{Daddi2009}, and our CO $1-0$ measurements.
\end{flushleft}
\end{table*}

\begin{figure}
\centering
 \includegraphics[scale=0.45]{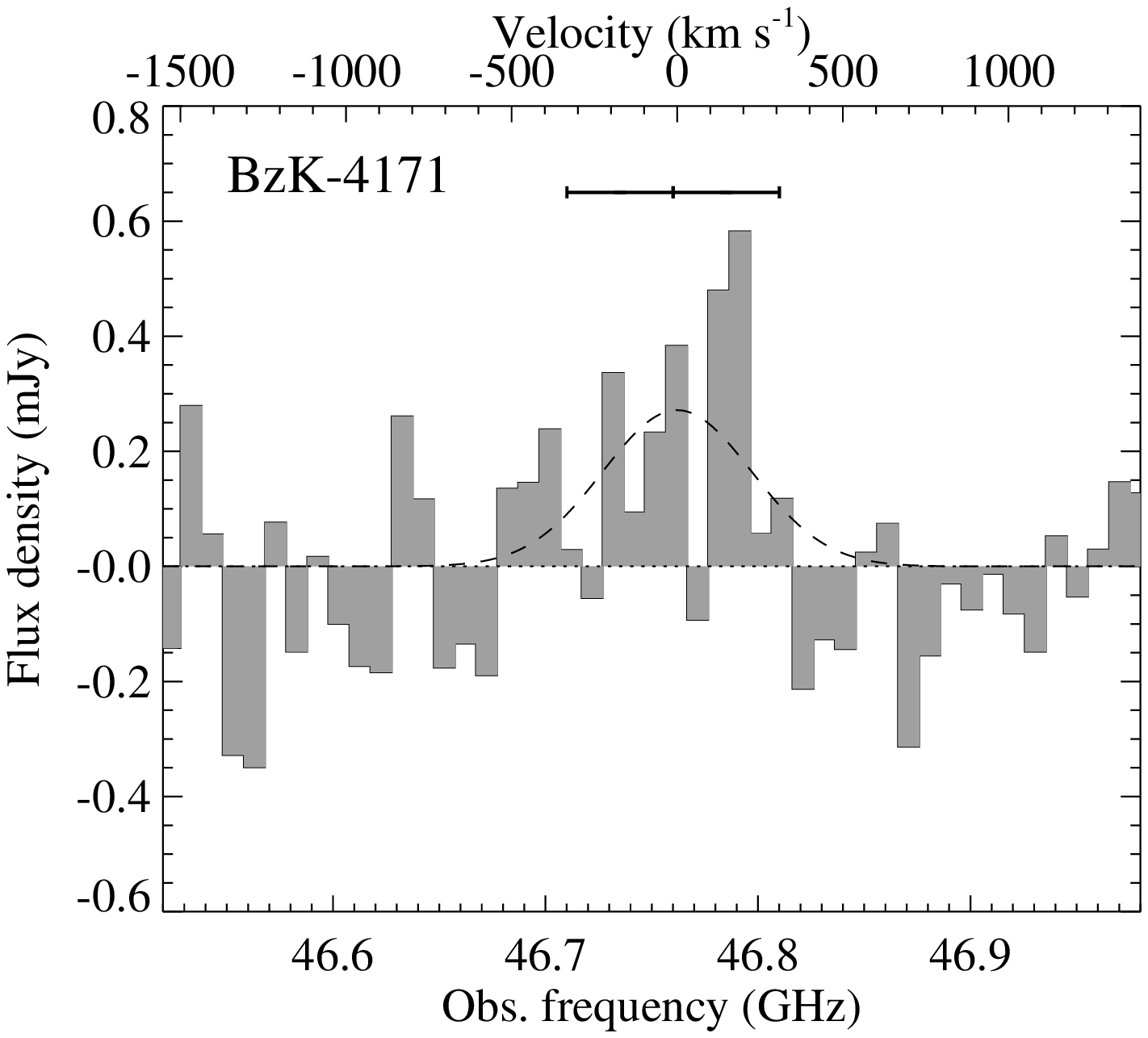}
 \includegraphics[scale=0.45]{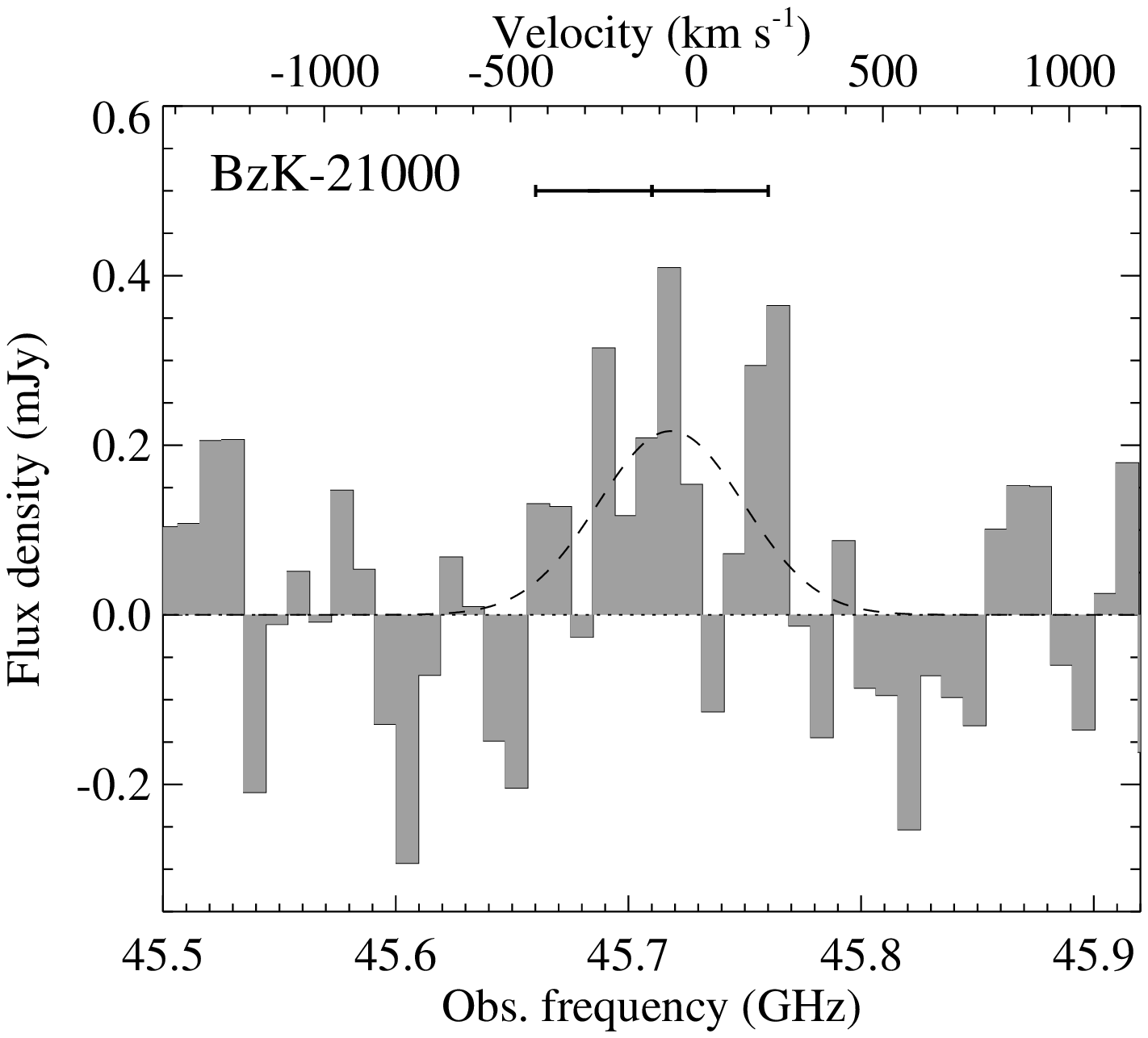}
  \caption{GBT spectra of BzK-4171 and BzK-21000. Horizontal marks indicate the position of the two 50 MHz channels observed by the VLA in each source. The line frequency obtained with the Gaussian fit (see text) was used to set the velocity scale. The dashed line shows single Gaussian fits to the spectra (see text), which were used to set the velocity scales.\label{fig:gbt_spectrum}}
\end{figure}

\subsection{Very Large Array}
We used the Very Large Array (VLA) in its C and D-array configuration and the Q-band receivers to observe the redshifted CO $1-0$ emission line ($\nu_\mathrm{rest}=115.271$ GHz) in three BzK galaxies at $z\sim1.5$ in the GOODS North field. This configuration provides good spatial resolution (typical beams of $\sim1-2\arcsec$), with a primary beam of $\sim 60\arcsec$. 

For two of our sources, BzK-4171 and BzK-21000, the observations were done in D-configuration between 2009 November 02 and 2009 November 27 under very good weather conditions. A total of 35 hrs were spent observing each source, over 5 tracks per source. Since the observations were done during a transition period to the Expanded VLA (EVLA), many antennae were malfunctioning and we were only able to reach less than half the sensitivity expected for the whole array (Table \ref{table:1}). 

We used two channels of 50 MHz bandwidth each and two polarizations per channel. At $\sim45$ GHz, 50 MHz correspond to $\sim330$ km s$^{-1}$ velocity coverage. For BzK-4171, the two channels were centered at 46.735 and 46.785 GHz. For BzK-21000, the two channels were centered at 45.685 GHz and 45.735 GHz. A fraction of the time of each track was spent observing these sources at 43 GHz using two non-overlapping channels of 50 MHz bandwidth (i.e. 100 MHz bandwidth in total) in order to obtain a limit for the continuum emission. In both cases the phase tracking center was pointed about 10\arcsec \ south from the target positions.  

The observations of BzK-16000 were done in C- and D-configuration between 2009 July and 2009 December under mostly good weather conditions. A total of 48 hours were used to observe this source. Spectral line observations were performed using two IFs of 7 channels each and two polarizations. Both IFs have one channel overlap, leading to a total of 13 independent spectral channels. These channels were combined into one single channel of 40.625 MHz bandwidth and centered at 45.653 GHz, covering most of the expected CO $1-0$ emission line. We also performed continuum observations of this source by placing two 50 MHz channels at each side of the expected CO $1-0$ line. These channels were centered at 45.585 GHz and 45.685 GHz, respectively. The higher frequency channel overlaps the central 40.625 MHz bandwidth region by 18 MHz.


In all cases, we used fast-switching calibration and observed the VLA calibrators J1302$+$5748 and 3C286 (J1331$+$305) for flux calibration. The AIPS software was used for data editing and calibration. Most of the data showed good phase stability; however, some time ranges were removed due to antennae with bad amplitudes. Finally, we used the AIPS task IMAGR, which employs the CLEAN algorithm, and natural weighting to deconvolve the images down to residuals of $\sim1\sigma$ in a box centered on our targets. Table \ref{table:1} summarizes the relevant observational parameters. 

\begin{figure*}
 \centering
\includegraphics[scale=0.38]{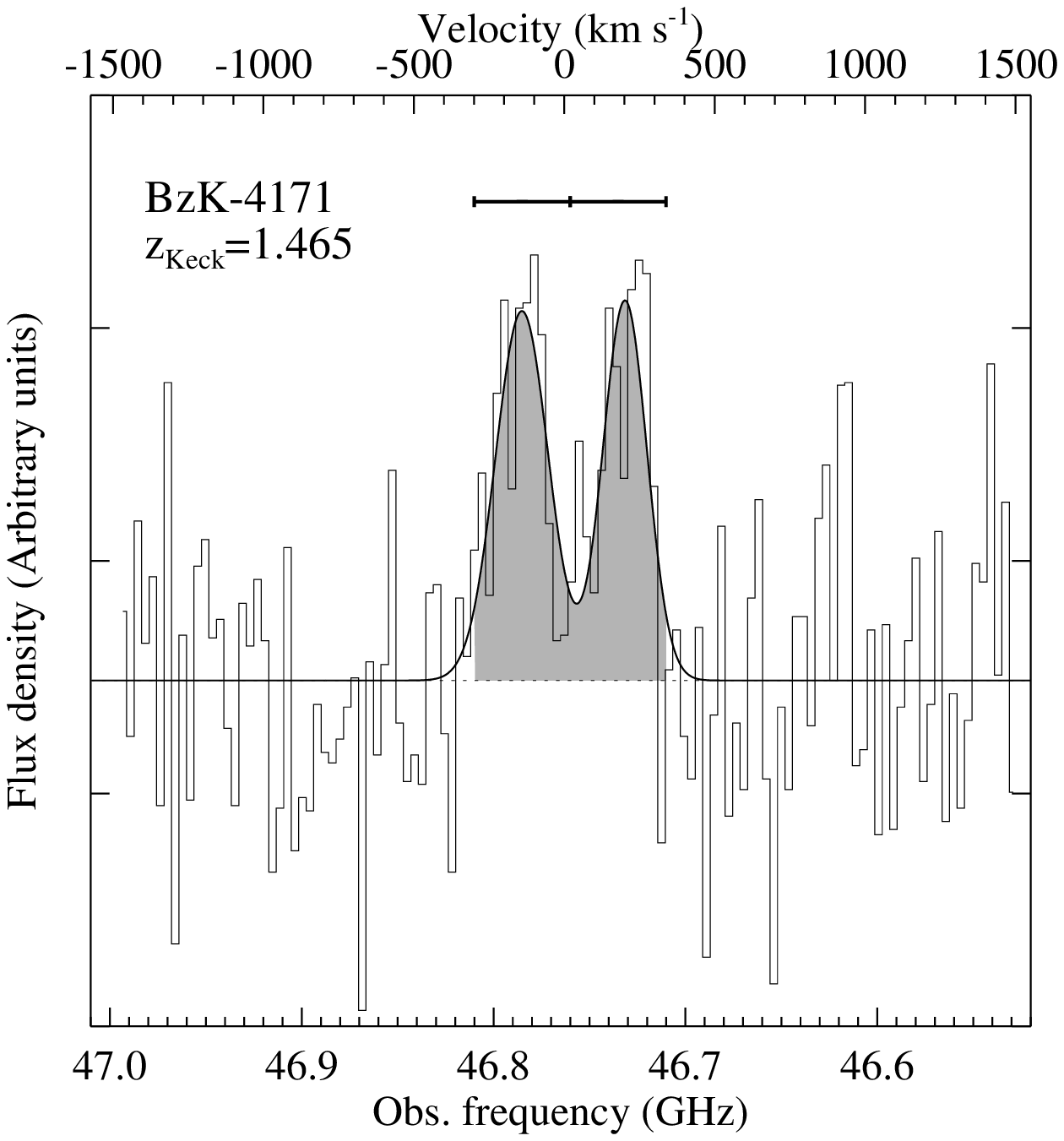}
\includegraphics[scale=0.38]{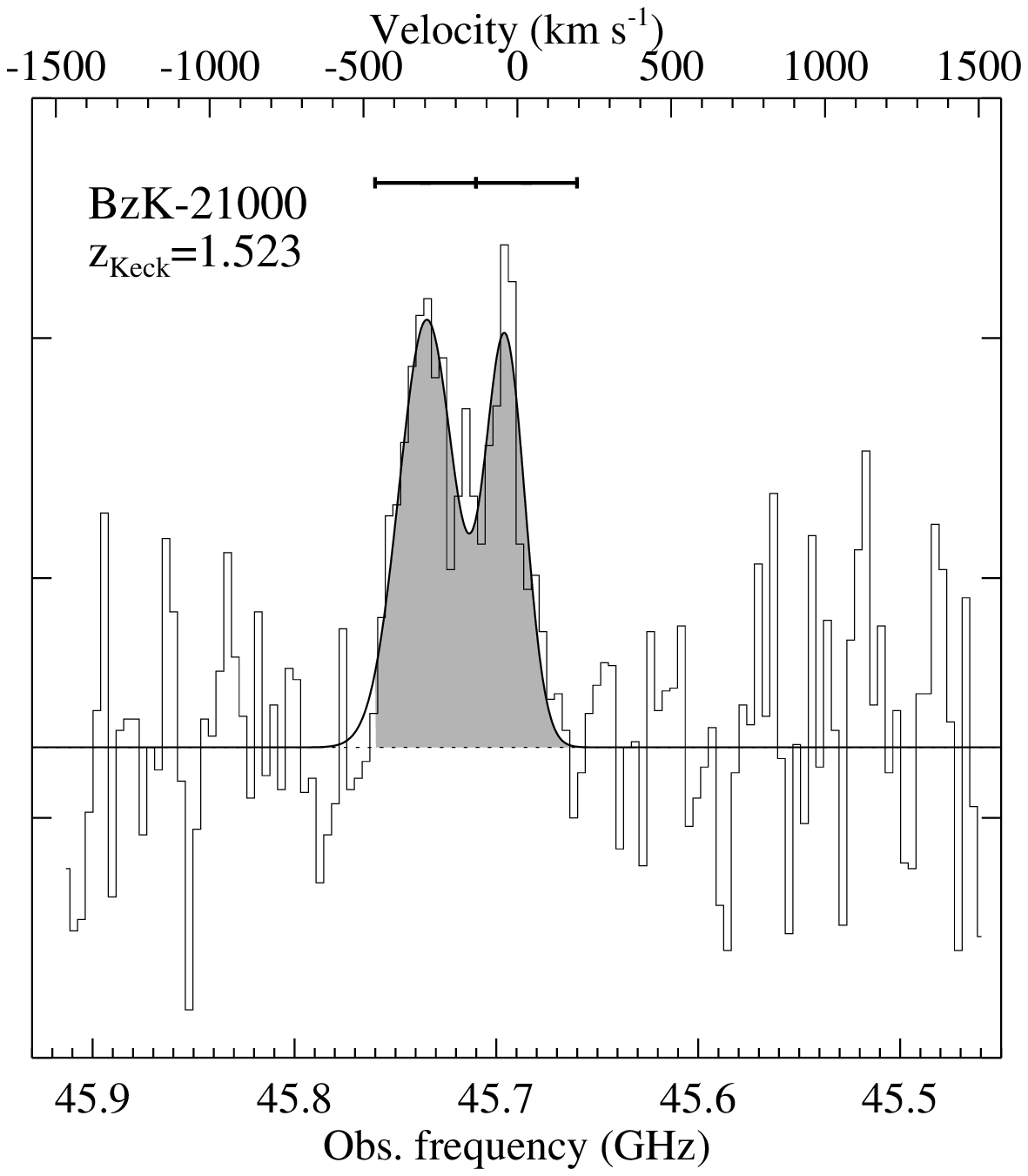}
\includegraphics[scale=0.38]{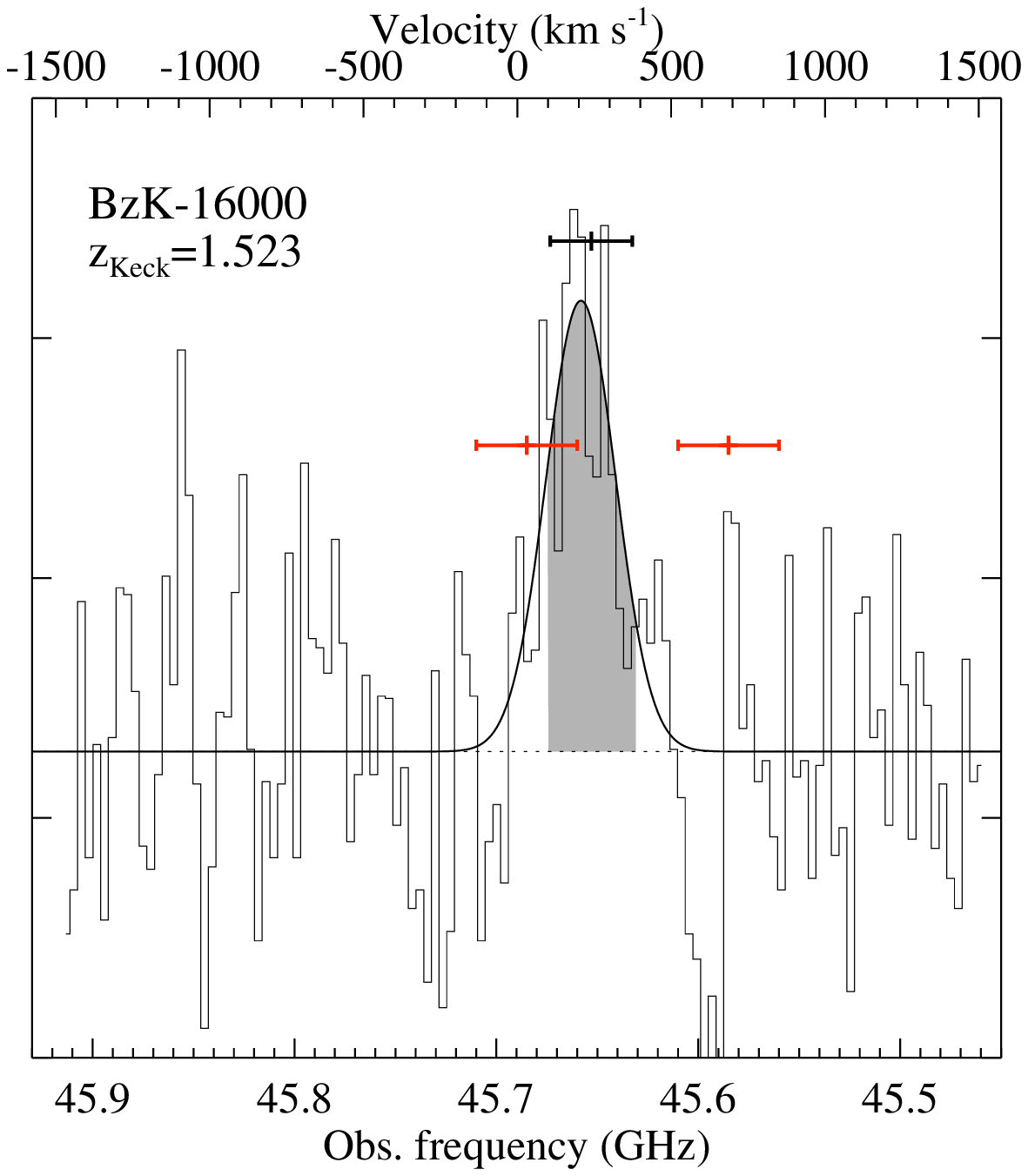}
\caption{PdBI CO $2-1$ spectra (in arbitrary flux units) of BzK-4171, BzK-21000 and BzK-16000 \citep[from][]{Daddi2009}, shifted to the frequency of the CO 1-0 transition to illustrate the fraction of the line covered by the VLA observations. A Gaussian fit to the emission is also shown. The bars on top as well as the shaded area in each spectrum indicates the spectral line coverage of the VLA bands. The lower red horizontal bars in the BzK-16000 panel show the two 50 MHz bands used to measure the continuum emission for that galaxy. \label{fig:coverage}}
\end{figure*}

\begin{figure*}
 \centering
\includegraphics[scale=0.34]{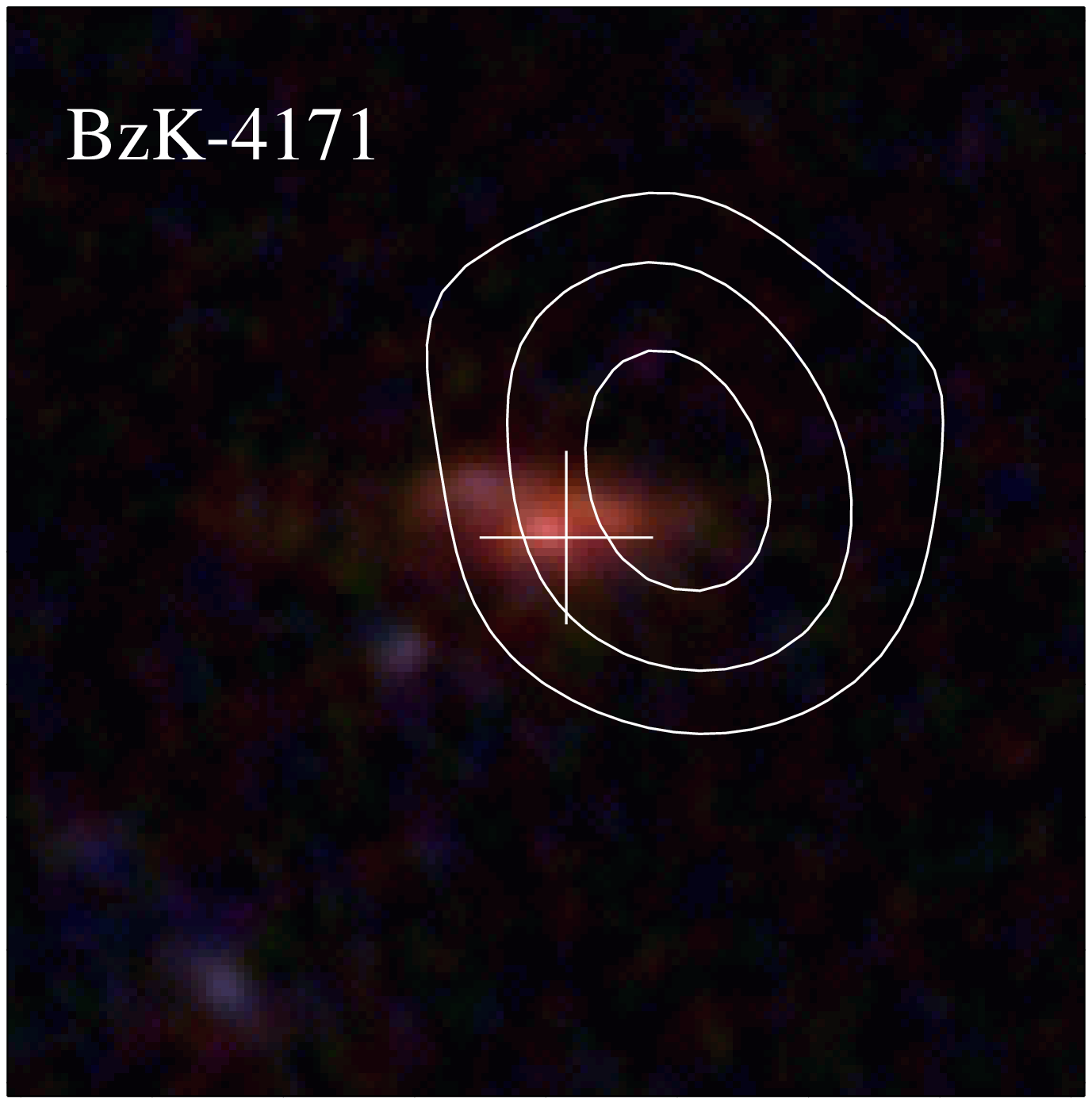}
\includegraphics[scale=0.34]{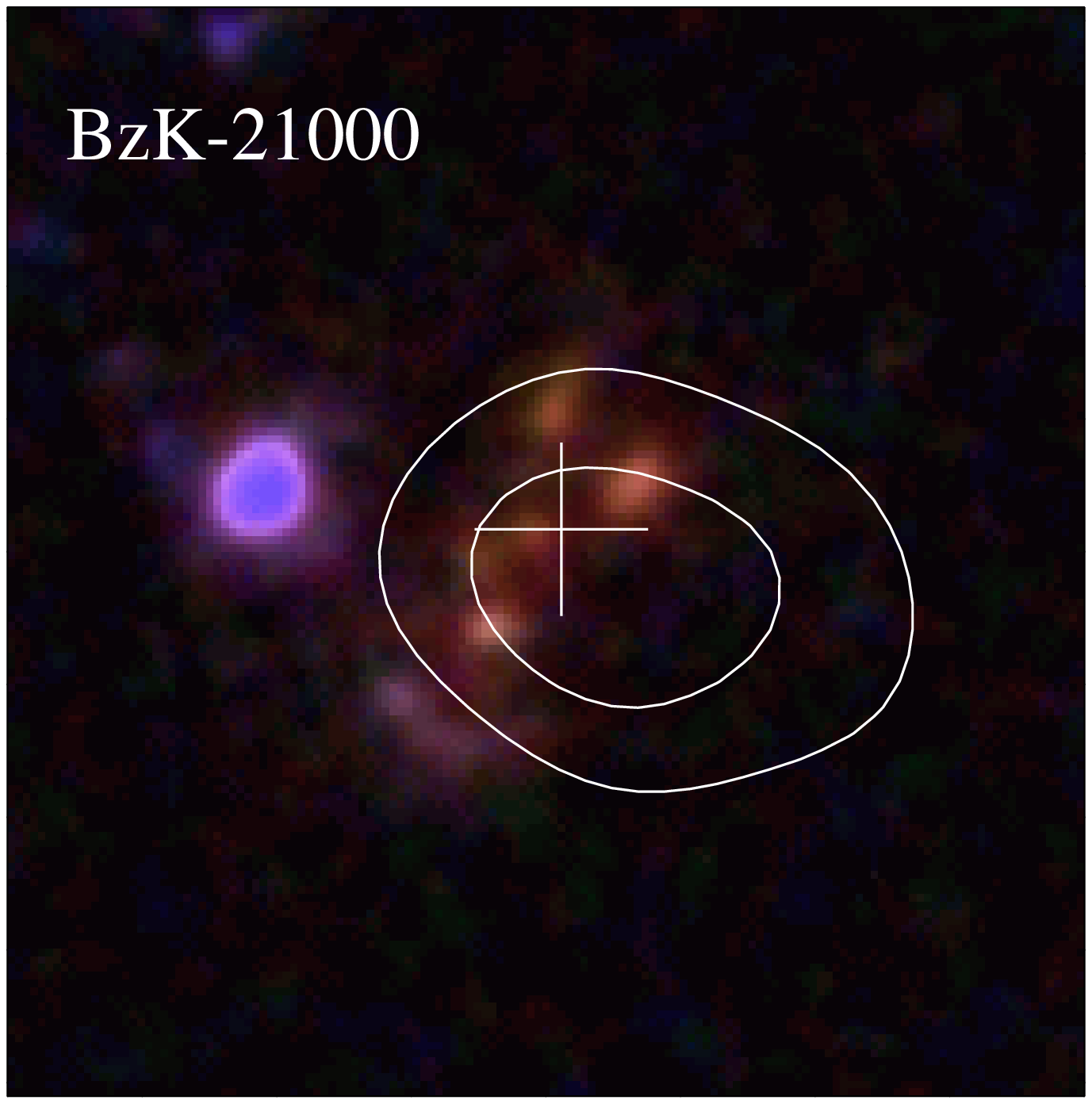}
\includegraphics[scale=0.34]{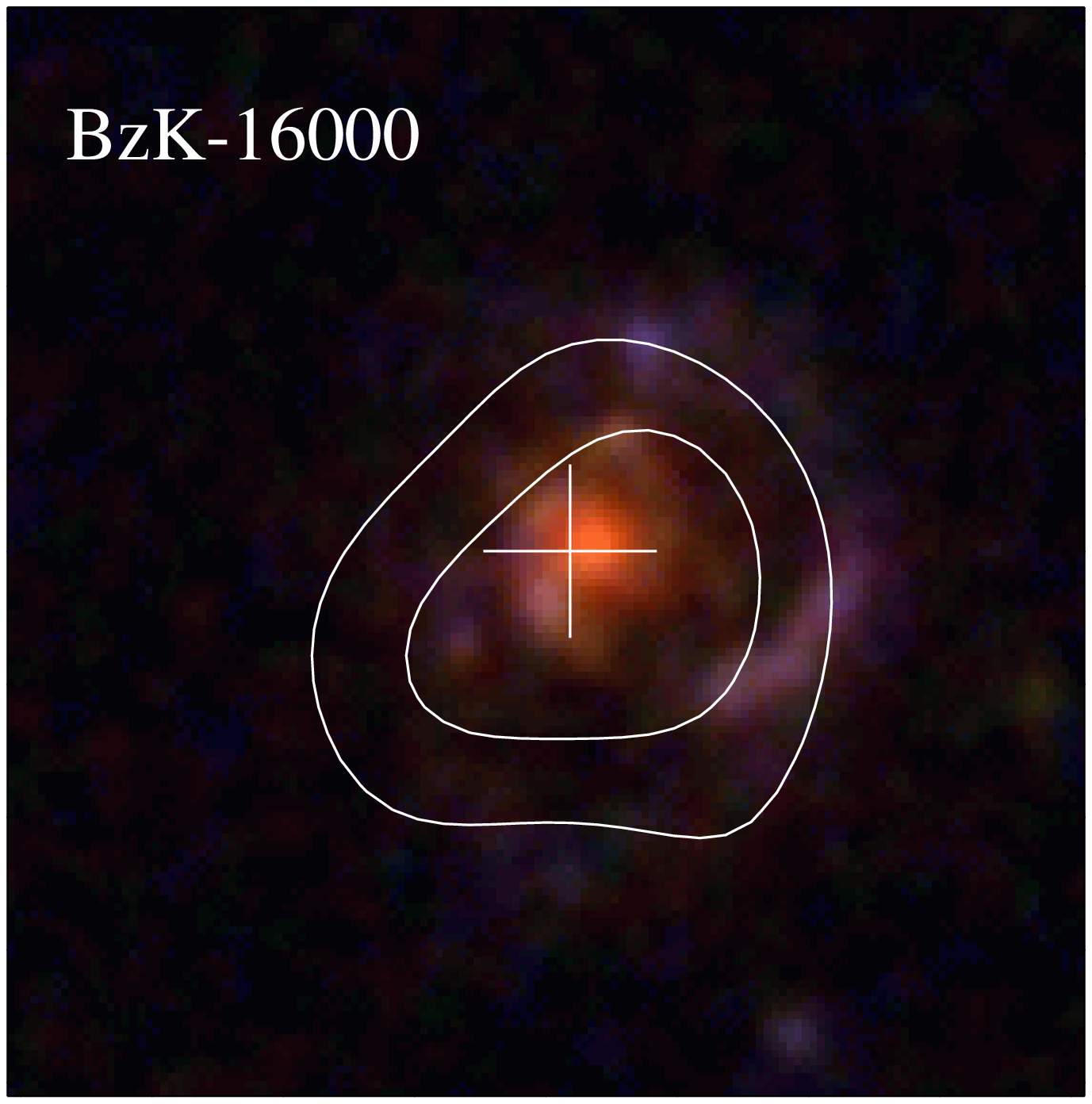}
\caption{VLA CO 1-0 emission line maps overlaid on the HST $BIz$ color images. Images are $4.2\arcsec\times4.2\arcsec$ in size. Contours and cross symbol are as in Fig. \ref{fig:bzk_im}. \label{fig:contours}}
\end{figure*}

\subsection{Green Bank Telescope}

We also performed observations of two of our BzK galaxies with the Robert C. Byrd Green Bank telescope (GBT) during 2009 April and 2009 October to November, under very good weather conditions. Typical computed opacities at 45 GHz are $\tau\sim0.1-0.2$ and measured wind velocities were $\lesssim2$ m s$^{-1}$. At these frequencies the GBT beam is $\sim16\arcsec$. We observed in sub-reflector nodding mode, with a half-cycle time of 6 s. At the beginning of each run, we observed the following flux density calibrators for 10 min each: 3C286, 3C147 or 3C295. We estimate our measured fluxes densities to be accurate within $\pm10\%$. We used the source 1259$+$514 as the pointing and focus calibrator. Pointing and focus were checked every 30 min to 1 h depending on the stability and accuracy of the obtained corrections in each run. Pointing was stable within $5\arcsec$ in all runs. Typical system temperatures at 45.7 GHz and 46.7 GHz were in the range 85-110 K. We employed two IFs of 800 MHz bandwidth each and two polarizations each, with a spectral resolution of 390.625 kHz per channel, or $\sim$2.6 km s$^{-1}$  per channel. We placed the center of both IFs 200 MHz from each other and about 100 MHz from the line frequency each. In this way, the overlap region covered a band of $\sim$600 MHz. 

The data were reduced using the GBTIDL software, removing a few scans that presented bad channels or strongly distorted baselines. For each source, we averaged all scans and both IFs. The combination of both IFs produce a gain of about $5-10\%$ in the obtained signal-to-noise ratio due to digitization noise. We fitted polynomials of order 3 and subtracted them from the averaged spectra. This eliminates baseline structure on scales $\sim2$ times larger than the expected linewidth of the CO lines. Finally, we downgraded their spectral resolution to 61 and 63 km s$^{-1}$ ($\sim9.5$ MHz) for BzK-21000 and BzK-4171, respectively.

\section{Results}
\label{section:res}

Figure \ref{fig:bzk_im} shows the VLA maps of CO $1-0$ line emission and Table \ref{table:properties} summarizes the results. Fitting two dimensional Gaussians to the CO images suggests marginally resolved sources, although the signal-to-noise is insufficient to provide accurate estimates of source sizes. In order to derive the total CO fluxes of the sources, we fit Gaussians that were constrained by the position and size of the sources derived from the CO $2-1$ observations at the Plateau de Bure Interferometer (PdBI) by \citet{Daddi2009}. Constraining the Gaussian fitting decreases the number of free parameters and permits a direct comparison of amplitudes, although it assumes that the CO $1-0$ and $2-1$ emission are spatially coincident. 

We find that all sources are detected with significances in the range $3-4\sigma$ in the CO maps, with no evidence for emission in the 43 GHz continuum maps of BzK-4171 and BzK-21000 nor in the 45 GHz continuum maps of BzK-16000. We derive 2$\sigma$ limits of 125, 144 and 360 $\mu$Jy for the continuum flux densities of BzK-4171, BzK-21000 and BzK-16000, respectively.


Figure  \ref{fig:gbt_spectrum} shows the GBT spectra. BzK-4171 and BzK-21000 are both marginally detected. Using a single Gaussian fit, we find peak flux densities of $S_{\mathrm{CO}\ 1-0}=0.30\pm0.14$ mJy and $S_{\mathrm{CO}\ 1-0}=0.22\pm0.10$ mJy, where the quoted errors account for a 10\% uncertainty in the flux calibration, linewidths of $430\pm190$ km s$^{-1}$ and $480\pm220$ km s$^{-1}$, and CO redshifts of $1.465\pm0.003$ and $1.521\pm0.003$ for BzK-4171 and BzK-21000, respectively. Our GBT measurements are consistent with those obtained with the VLA in terms of flux densities, and with those obtained with the PdBI for the CO $2-1$ line in terms of linewidths and central frequencies. Given the lower S/N of the GBT observations, we refer to the VLA flux measurements in the remainder of this paper.   

Figure \ref{fig:coverage} shows the PdBI CO $2-1$ spectra of our sources \citep{Daddi2009} scaled to the frequency of the CO $1-0$ transition. The horizontal bar and shaded region of the emission lines represent the velocity (and/or frequency) range covered by our VLA CO $1-0$ observations. The velocity ranges covered are 641 km s$^{-1}$, 656 km s$^{-1}$ and 266 km s$^{-1}$ in BzK-4171, BzK-21000 and BzK-16000, respectively. This implies that we cover $96\%$, $98\%$ and $76\%$, respectively, of the CO $1-0$ line with our observations (Table \ref{table:1}). After correcting for the fraction of the line that falls outside our bands, we derive integrated line intensities, $I_\mathrm{CO}=\int S_\mathrm{CO} dv$, of $0.20\pm0.05$, $0.13\pm0.03$ and $0.28\pm0.08$ Jy km s$^{-1}$ in BzK-4171, BzK-21000 and BzK-16000, respectively. We use the integrated line intensities to compute the CO line luminosities through $L_\mathrm{CO}'=3.25\times10^7 (1+z)^{-1} D_\mathrm{L}^{2} \nu_\mathrm{obs}^{-2} I_\mathrm{CO}$, where $D_\mathrm{L}$ is the luminosity distance at redshift $z$ and $\nu_\mathrm{obs}$ is the observed frequency \citep{Solomon1997}. 

A comparison of the CO $1-0$ maps with optical and the radio positions (Fig. \ref{fig:contours}) shows a small offset of $\sim0.3\arcsec$ from the CO peak position in the case of BzK-4171. This is well within the range expected given the low significance of the detection. For the other sources, the CO emission is consistent ($<0.2\arcsec$) with the position of the radio and optical source.

\begin{figure}
 \centering
\includegraphics[scale=0.5]{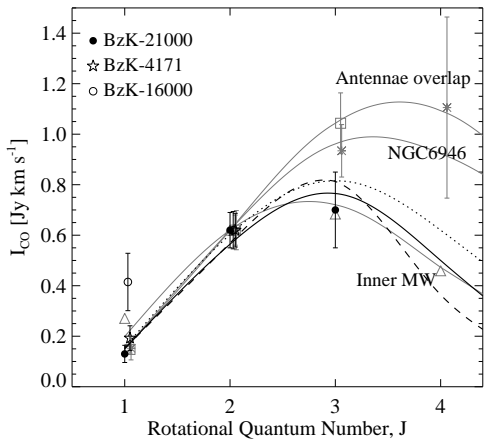}
\caption{CO excitation ladder or CO SED for BzK-21000 (solid black circles). The black solid, dashed and dotted lines show representative LVG models 1, 2 and 3 from \citet{Dannerbauer2009}, respectively. The open stars and open circles show the CO $1-0$ and CO $2-1$ integrated flux densities for BzK-4171 and BzK-16000, scaled to the CO $2-1$ emission of BzK-21000 for comparison. The open triangle, open square and asterisk symbols illustrate the CO SED of the Milky Way, the Antennae overlap region and the spiral galaxy NGC6946, normalized to the CO $2-1$ emission of BzK-21000. The LVG models that reproduce the data for these objects are shown as gray solid lines. A small horizontal shift has been applied to the data points to enhance the visibility.\label{fig:lvg}}
\end{figure}

\subsection{Gas properties}


Our observations of the CO $1-0$ line emission can be used to directly measure the amount of molecular gas and star formation efficiency in the BzK galaxies. The estimation of the molecular gas masses is typically done by using a CO luminosity to gas mass conversion factor, $\alpha_\mathrm{CO}$. Comparing observations with simulations of star-forming disk galaxies that reproduced the observed CO $2-1$ line shapes and UV morphologies of three BzK galaxies (two of which are in our sample), \citet{Daddi2009} estimated $\alpha_\mathrm{CO}=3.6\pm0.8\ M_\sun$ (K km s$^{-1}$ pc$^{2}$)$^{-1}$, close to the Galactic disk $\alpha$. Assuming this value of $\alpha_\mathrm{CO}$, we compute the H$_2$ masses given in Table \ref{table:properties}. 

The star formation efficiency of galaxies is defined as the ratio between the IR luminosity and the CO $1-0$ luminosity, SFE$=L_\mathrm{IR}/L_\mathrm{CO}'$. Using the values for the IR luminosity from \citet{Daddi2009}, we find SFE$=43,\ 142$ and 44 $L_\sun$ (K km s$^{-1}$ pc$^2$)$^{-1}$ for BzK-4171, BzK-21000 and BzK-16000, respectively. We can also compute the time in which the gas would be consumed if the current SFR remains constant, $\tau_\mathrm{gas}=M(\mathrm{H}_2)/$SFR. Using the SFRs found by \citet{Daddi2009}, we find gas consumption lifetimes of $\sim0.2-0.8$ Gyr (see Table \ref{table:properties}). These values are comparable to those found by \citet{Daddi2009} and are similar those found for comparable galaxies at redshifts $\sim1-3$ \citep{Tacconi2010}.

\subsection{Excitation of the molecular gas}

In this section we study the properties of the molecular gas in our BzK galaxies. All of our sources have previously been detected in the CO $2-1$ emission line. However, only BzK-21000 has been detected in the CO $3-2$ line \citep{Dannerbauer2009}.

Figure \ref{fig:lvg} shows the velocity integrated CO line fluxes of BzK-21000 as a function of rotational quantum number, $J$. Also shown are the CO $1-0$ and CO $2-1$ integrated line fluxes of BzK-4171 and BzK-16000, normalized to the CO $2-1$ intensity of BzK-21000 for comparison, and the normalized CO intensities for the inner disk of the Milky Way \citep{Fixsen1999}, the Antenna overlap region \citep{Zhu2003} and the spiral galaxy NGC6946 \citep{Bayet2006}. 

From the velocity integrated line fluxes, we compute the brightness temperature line ratios as $r_{21}=T_{21}/T_{10}=(I_{21})/I_{10})\times(\nu_{10}/\nu_{21})^2$, where $T_{21}$ and $T_{10}$ are the brightness temperatures, $I_{21}$ and $I_{10}$ are the integrated fluxes, and $\nu_{21}$ and $\nu_{10}$ are the observed frequencies of the CO $2-1$ and $1-0$ emission lines, respectively. For LTE, we expect this ratio to be $r_{21}=1$. In our sources, we measure brightness temperature line ratios of $0.80_{-0.22}^{+0.35}$, $1.22_{-0.36}^{+0.61}$ and $0.41_{-0.13}^{+0.23}$ for BzK-4171, BzK-21000 and BzK-16000, respectively. 

At the significance of our detections, it is not possible to conclude whether the emission in the individual BzK galaxies is thermalized or not up to $J=2$. In BzK-4171 and BzK-21000 the values of $r_{21}$ suggest thermal equilibrium, while in BzK-16000, the low value for $r_{21}$ suggests  that the CO emission is sub-thermal at $J=2$. However, this needs to be confirmed with deeper observations. For BzK-21000, the brightness temperature ratio between CO $3-2$ and CO $1-0$ is $r_{31}=0.61_{-0.26}^{0.39}$, compatible with the previous results of $r_{31}\sim0.5$ in this galaxy by \citet{Dannerbauer2009}.


For BzK-21000, we compute a Large Velocity Gradient (LVG) model using our new CO $1-0$ measurement. We employ a single component LVG model that assumes spherical geometry. We use the collision rates from \citet{Flower2001} with an ortho-para H$_2$ ratio of 3 and a CO abundance per velocity gradient [CO]/($dv/dr$)=$10^{-5}$ pc (km s$^{-1}$)$^{-1}$ \citep[e.g.,][]{Weiss2005b, Weiss2007}. Values that resemble the CO emission are in the range $T_\mathrm{kin}=20-150$ K and $n(\mathrm{H}_2)=400-2500$ cm$^{-3}$. Since we do not have sufficient constraints to fit the data with a specific model, we thus discuss the three representative models presented by \citet{Dannerbauer2009}, as shown in Fig. \ref{fig:lvg}. These models have $T_\mathrm{kin}=25$ K, $n(\mathrm{H}_2)=1300$ cm$^{-3}$ and a cloud filling factor of $\sim$2\% (Model 1); $T_\mathrm{kin}=90$ K, $n(\mathrm{H}_2)=600$ cm$^{-3}$ and a similar cloud filling factor (Model 2); and $T_\mathrm{kin}=10$ K, $n(\mathrm{H}_2)=2500$ cm$^{-3}$ and a  filling factor of 8\% (Model 3). We see that all these models can reasonably reproduce the data for BzK-21000. This is expected as our CO $1-0$ flux measurement confirms the LVG-based prediction of  \citet{Dannerbauer2009}, $I_{\mathrm{CO}\ 1-0}\sim0.15$ Jy km s$^{-1}$. Therefore, we verify that models with high $T_\mathrm{kin}$ and/or $n(\mathrm{H}_2)$, as seen in ULIRGs or high-redshift QSOs, are unlikely for the gas in this galaxy, as suggested by \citet{Dannerbauer2009}.

\section{Summary and Discussion}
\label{section:diss}

\begin{figure}
 \centering
\includegraphics[scale=0.5]{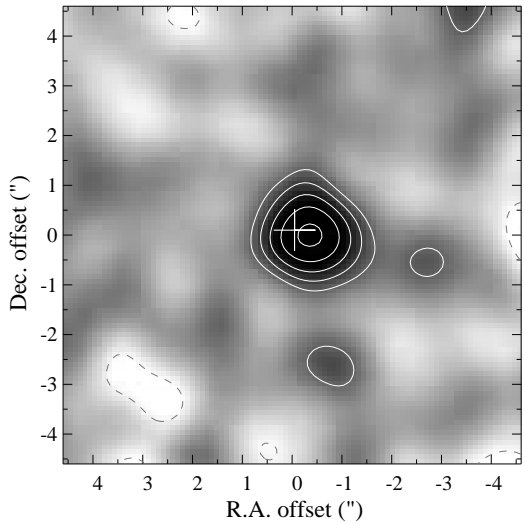}
\caption{Combined CO $1-0$ emission from the three BzK galaxies. Contours levels are: -2, 2, 3, 4, 5 and 6$\sigma$. The cross indicates the reference VLA position.\label{fig:stack}}
\end{figure}

We have detected the CO $1-0$ emission line in three massive star-forming galaxies at $z\sim1.5$. Our observations allow us to carry out a direct comparison with studies of galaxies at low-redshift, and hence put unique constraints on the properties of the molecular gas in disk galaxies at high-redshift. 
 
From our molecular gas excitation analysis for BzK-21000, we confirm the \citet{Dannerbauer2009} results that the gas in this galaxy has low excitation conditions at $J=3$ (Fig. \ref{fig:lvg}). The CO ladder for this galaxy seems to be thermalized up to $J=2$ and it is sub-thermal at $J=3$. The CO ladder is similar to that found for local disks, as in NGC6946 \citep[Fig. \ref{fig:lvg};][]{Mauersberger1999}. However, at the significance of this detection, the CO ladder is also consistent with the Milky Way galaxy, for which the emission appears to be non-LTE already at $J=2$.

A similar case is that of BzK-4171, where the relatively large uncertainty in the temperature ratio, $r_{21}$, does not allow us to differentiate between thermal or sub-thermal molecular gas up to $J=2$. Measurements of the CO $3-2$ transition are necessary to check whether or not higher order CO transitions are in LTE in this galaxy. Interestingly, in the case of BzK-16000, our measurements suggest that the CO emission is sub-thermal ($r_{21}<1$) already at $J=2$, resembling what appears to be the case for the inner disk of the Milky Way galaxy \citep{Fixsen1999}. Our estimate of $r_{21}\sim0.4$ differs from the LTE at the $\approx2\sigma$ level. Measurements of higher$-J$ transitions will help to study this particular galaxy in more detail.

Based on the previous CO $1-0$ observations of BzK-21000 \citep{Dannerbauer2009}, \citet{Daddi2009} and \citet{Tacconi2010} assumed $r_{31}=0.5$ to convert their CO $3-2$ luminosities into CO $1-0$ luminosities. For this galaxy, we obtain $r_{31}\sim0.6$, which validates, within $1\sigma$, their assumption.

For a solid measurement of the average CO $1-0$ line emission and the average molecular gas properties of our BzK galaxies, we extracted $10\arcsec\times10\arcsec$ cutouts centered at the VLA peak positions and produced a noise-weighted average CO $1-0$ map of their combined emission (Fig. \ref{fig:stack}). We obtain a detection with a flux density of $225\pm36\ \mu$Jy ($6.25\sigma$). Stacking with respect to the CO position yields a flux density of $220\pm40\ \mu$Jy (5.5$\sigma$), consistent within the uncertainties ($<1\sigma$) with the flux measured by stacking with respect to the VLA position. From the values presented in \citet{Daddi2009} for the CO $2-1$ emission, we compute a noise-weighted average CO $2-1$ to CO $1-0$ brightness temperature ratio of $0.92_{-0.19}^{+0.28}$, suggesting that the CO SED from these galaxies is thermalized up to $J=2$. This ratio is consistent with what is found in local disk galaxies, which have lower CO luminosities than our BzK galaxies \citep[][ and references therein]{Mauersberger1999, Bayet2006}, but it is also similar to other galaxy populations with CO luminosities comparable to that of our BzK galaxies, such as local ULIRGs \citep{Weiss2005b, Guesten2006, Hitschfeld2008, Bayet2006, Bayet2009}, and high-redshift SMGs and QSOs \citep{Solomon2005, Weiss2005b, Riechers2006, Weiss2007}. Thus, our comparison based only on CO $2-1$ to CO $1-0$ ratios is not yet sufficient to decide whether the excitation conditions in our BzK galaxies are, on average, different or not from those of other galaxy populations. 

We note that \citet{Daddi2009} assumed $r_{21}=0.86$ to convert the CO $2-1$ luminosities into CO $1-0$ luminosities for their sample of 6 BzK galaxies, following the \citet{Dannerbauer2009} results on BzK-21000. This assumption is supported by our measurement from the average stacked CO map.

From our stacked CO $1-0$ measurement, we find on average a SFE $\approx100$ (K km s$^{-1}$ pc$^2$)$^{-1}$. This, including individual values, is well below the average for high-redshift SMGs, $560\pm210$ $L_\sun$ (K km s$^{-1}$ pc$^2$)$^{-1}$ \citep{Greve2005, Tacconi2006} and similar to the values found for local spiral galaxies \citep{Boselli2002, Leroy2008} and for other disk galaxies at high-redshift \citep{Tacconi2010}. However, it is marginally consistent with the average value found in local ULIRGs, $\sim225$ $L_\sun$ (K km s$^{-1}$ pc$^2$)$^{-1}$ \citep{Solomon1997, Yao2003}. We obtain an average gas depletion lifetime of $\sim0.4$ Gyr for the galaxies in our sample, which is higher than what is found in SMGs, $<0.05$ Gyr \citep{Tacconi2006}. 

Observations of disk galaxies at high-redshift (including this work) have provided detections only for the lower-$J$ CO transitions, and thus they probe the lower end of the CO SED. Recent studies of high-redshift SMGs and QSOs have indicated that two gas components can be present: a cold, low density gas component and a warm, dense component \citep[e.g., ][]{Weiss2007, Carilli2010}. If indeed such is the case for star-forming BzK galaxies, observations of higher-$J$ CO lines are crucial. High-definition CO multi-transition studies over larger samples are necessary. Here, the EVLA and Atacama Large Millimeter Array (ALMA) will play a fundamental role, disentangling the molecular gas distribution of the cold molecular gas traced by CO $1-0$ and the warm gas probed by the higher-$J$ transitions.

\acknowledgments
MA thanks A. Leroy for useful discussions. We thank Christian Henkel for providing us with the Large Velocity Gradient code in its original version. CC thanks the Max-Planck-Gesellschaft and the Humboldt-Stiftung for support through the Max-Planck-Forschungspreis. DR acknowledges support from from NASA through Hubble Fellowship grant HST-HF-51235.01 awarded by the Space Telescope Science Institute, which is operated by the Association of Universities for Research in Astronomy, Inc., for NASA, under contract NAS 5-26555. The work of DS was carried out at Jet Propulsion Laboratory, California Institute of Technology, under a contract with NASA.

\bibliographystyle{apj}
\bibliography{manuel_bzk}

\begin{thebibliography}{41}
\expandafter\ifx\csname natexlab\endcsname\relax\def\natexlab#1{#1}\fi

\bibitem[{{Aravena} {et~al.}(2008){Aravena}, {Bertoldi}, {Schinnerer}, {Weiss},
  {Jahnke}, {Carilli}, {Frayer}, {Henkel}, {Brusa}, {Menten}, {Salvato}, \&
  {Smolcic}}]{Aravena2008}
{Aravena}, M., {et~al.} 2008, \aap, 491, 173

\bibitem[{{Bayet} {et~al.}(2006){Bayet}, {Gerin}, {Phillips}, \&
  {Contursi}}]{Bayet2006}
{Bayet}, E., {Gerin}, M., {Phillips}, T.~G., \& {Contursi}, A. 2006, \aap, 460,
  467

\bibitem[{{Bayet} {et~al.}(2009){Bayet}, {Gerin}, {Phillips}, \&
  {Contursi}}]{Bayet2009}
---. 2009, \mnras, 399, 264

\bibitem[{{Bell} {et~al.}(2005){Bell}, {Papovich}, {Wolf}, {Le Floc'h},
  {Caldwell}, {Barden}, {Egami}, {McIntosh}, {Meisenheimer},
  {P{\'e}rez-Gonz{\'a}lez}, {Rieke}, {Rieke}, {Rigby}, \& {Rix}}]{Bell2005}
{Bell}, E.~F., {et~al.} 2005, \apj, 625, 23

\bibitem[{{Boselli} {et~al.}(2002){Boselli}, {Lequeux}, \&
  {Gavazzi}}]{Boselli2002}
{Boselli}, A., {Lequeux}, J., \& {Gavazzi}, G. 2002, \apss, 281, 127

\bibitem[{{Carilli} {et~al.}(2007){Carilli}, {Neri}, {Wang}, {Cox}, {Bertoldi},
  {Walter}, {Fan}, {Menten}, {Wagg}, {Maiolino}, {Omont}, {Strauss},
  {Riechers}, {Lo}, {Bolatto}, \& {Scoville}}]{Carilli2007}
{Carilli}, C.~L., {et~al.} 2007, \apjl, 666, L9

\bibitem[{{Carilli} {et~al.}(2010){Carilli}, {Daddi}, {Riechers}, {Walter},
  {Weiss}, {Dannerbauer}, {Morrison}, {Wagg}, {Dav{\'e}}, {Elbaz}, {Stern},
  {Dickinson}, {Krips}, \& {Aravena}}]{Carilli2010}
---. 2010, \apj, 714, 1407

\bibitem[{{Coppin} {et~al.}(2008){Coppin}, {Swinbank}, {Neri}, {Cox},
  {Alexander}, {Smail}, {Page}, {Stevens}, {Knudsen}, {Ivison}, {Beelen},
  {Bertoldi}, \& {Omont}}]{Coppin2008}
{Coppin}, K.~E.~K., {et~al.} 2008, \mnras, 389, 45

\bibitem[{{Daddi} {et~al.}(2008){Daddi}, {Dannerbauer}, {Elbaz}, {Dickinson},
  {Morrison}, {Stern}, \& {Ravindranath}}]{Daddi2008}
{Daddi}, E., {Dannerbauer}, H., {Elbaz}, D., {Dickinson}, M., {Morrison}, G.,
  {Stern}, D., \& {Ravindranath}, S. 2008, \apjl, 673, L21


\bibitem[{{Daddi} {et~al.}(2010{\natexlab{a}}){Daddi}, {Bournaud}, {Walter},
  {Dannerbauer}, {Carilli}, {Dickinson}, {Elbaz}, {Morrison}, {Riechers},
  {Onodera}, {Salmi}, {Krips}, \& {Stern}}]{Daddi2009}
{Daddi}, E., {et~al.} 2010{\natexlab{a}}, \apj, 713, 686

\bibitem[{{Daddi} {et~al.}(2010{\natexlab{b}}){Daddi}, {Elbaz}, {Walter},
  {Bournaud}, {Salmi}, {Carilli}, {Dannerbauer}, {Dickinson}, {Monaco}, \&
  {Riechers}}]{Daddi2010b}
---. 2010{\natexlab{b}}, \apjl, 714, L118

\bibitem[{{Dannerbauer} {et~al.}(2009){Dannerbauer}, {Daddi}, {Riechers},
  {Walter}, {Carilli}, {Dickinson}, {Elbaz}, \& {Morrison}}]{Dannerbauer2009}
{Dannerbauer}, H., {Daddi}, E., {Riechers}, D.~A., {Walter}, F., {Carilli},
  C.~L., {Dickinson}, M., {Elbaz}, D., \& {Morrison}, G.~E. 2009, \apjl, 698,
  L178

\bibitem[{{Dekel} {et~al.}(2009){Dekel}, {Birnboim}, {Engel}, {Freundlich},
  {Goerdt}, {Mumcuoglu}, {Neistein}, {Pichon}, {Teyssier}, \&
  {Zinger}}]{Dekel2009}
{Dekel}, A., {et~al.} 2009, \nat, 457, 451

\bibitem[{{Elbaz} {et~al.}(2007){Elbaz}, {Daddi}, {Le Borgne}, {Dickinson},
  {Alexander}, {Chary}, {Starck}, {Brandt}, {Kitzbichler}, {MacDonald},
  {Nonino}, {Popesso}, {Stern}, \& {Vanzella}}]{Elbaz2007}
{Elbaz}, D., {et~al.} 2007, \aap, 468, 33

\bibitem[{{Fixsen} {et~al.}(1999){Fixsen}, {Bennett}, \& {Mather}}]{Fixsen1999}
{Fixsen}, D.~J., {Bennett}, C.~L., \& {Mather}, J.~C. 1999, \apj, 526, 207

\bibitem[{{Flower}(2001)}]{Flower2001}
{Flower}, D.~R. 2001, Journal of Physics B Atomic Molecular Physics, 34, 2731

\bibitem[{{Genzel} {et~al.}(2008){Genzel}, {Burkert}, {Bouch{\'e}}, {Cresci},
  {F{\"o}rster Schreiber}, {Shapley}, {Shapiro}, {Tacconi}, {Buschkamp},
  {Cimatti}, {Daddi}, {Davies}, {Eisenhauer}, {Erb}, {Genel}, {Gerhard},
  {Hicks}, {Lutz}, {Naab}, {Ott}, {Rabien}, {Renzini}, {Steidel}, {Sternberg},
  \& {Lilly}}]{Genzel2008}
{Genzel}, R., {et~al.} 2008, \apj, 687, 59

\bibitem[{{Greve} {et~al.}(2005){Greve}, {Bertoldi}, {Smail}, {Neri},
  {Chapman}, {Blain}, {Ivison}, {Genzel}, {Omont}, {Cox}, {Tacconi}, \&
  {Kneib}}]{Greve2005}
{Greve}, T.~R., {et~al.} 2005, \mnras, 359, 1165

\bibitem[{{G{\"u}sten} {et~al.}(2006){G{\"u}sten}, {Philipp}, {Wei{\ss}}, \&
  {Klein}}]{Guesten2006}
{G{\"u}sten}, R., {Philipp}, S.~D., {Wei{\ss}}, A., \& {Klein}, B. 2006, \aap,
  454, L115

\bibitem[{{Hitschfeld} {et~al.}(2008){Hitschfeld}, {Aravena}, {Kramer},
  {Bertoldi}, {Stutzki}, {Bensch}, {Bronfman}, {Cubick}, {Fujishita}, {Fukui},
  {Graf}, {Honingh}, {Ito}, {Jakob}, {Jacobs}, {Klein}, {Koo}, {May}, {Miller},
  {Miyamoto}, {Mizuno}, {Onishi}, {Park}, {Pineda}, {Rabanus}, {R{\"o}llig},
  {Sasago}, {Schieder}, {Simon}, {Sun}, {Volgenau}, {Yamamoto}, \&
  {Yonekura}}]{Hitschfeld2008}
{Hitschfeld}, M., {et~al.} 2008, \aap, 479, 75

\bibitem[{{Kere{\v s}} {et~al.}(2005){Kere{\v s}}, {Katz}, {Weinberg}, \&
  {Dav{\'e}}}]{Keres2005}
{Kere{\v s}}, D., {Katz}, N., {Weinberg}, D.~H., \& {Dav{\'e}}, R. 2005,
  \mnras, 363, 2

\bibitem[{{Leroy} {et~al.}(2008){Leroy}, {Walter}, {Brinks}, {Bigiel}, {de
  Blok}, {Madore}, \& {Thornley}}]{Leroy2008}
{Leroy}, A.~K., {Walter}, F., {Brinks}, E., {Bigiel}, F., {de Blok}, W.~J.~G.,
  {Madore}, B., \& {Thornley}, M.~D. 2008, \aj, 136, 2782

\bibitem[{{Lilly} {et~al.}(1996){Lilly}, {Le Fevre}, {Hammer}, \&
  {Crampton}}]{Lilly1996}
{Lilly}, S.~J., {Le Fevre}, O., {Hammer}, F., \& {Crampton}, D. 1996, \apjl,
  460, L1+

\bibitem[{{Madau} {et~al.}(1996){Madau}, {Ferguson}, {Dickinson}, {Giavalisco},
  {Steidel}, \& {Fruchter}}]{Madau1996}
{Madau}, P., {Ferguson}, H.~C., {Dickinson}, M.~E., {Giavalisco}, M.,
  {Steidel}, C.~C., \& {Fruchter}, A. 1996, \mnras, 283, 1388

\bibitem[{{Mauersberger} {et~al.}(1999){Mauersberger}, {Henkel}, {Walsh}, \&
  {Schulz}}]{Mauersberger1999}
{Mauersberger}, R., {Henkel}, C., {Walsh}, W., \& {Schulz}, A. 1999, \aap, 341,
  256

\bibitem[{{Morrison} {et~al.}(2010){Morrison}, {Owen}, {Dickinson}, {Ivison},
  \& {Ibar}}]{Morrison2010}
{Morrison}, G.~E., {Owen}, F.~N., {Dickinson}, M., {Ivison}, R.~J., \& {Ibar},
  E. 2010, \apjs, 188, 178

\bibitem[{{Riechers} {et~al.}(2008){Riechers}, {Walter}, {Brewer}, {Carilli},
  {Lewis}, {Bertoldi}, \& {Cox}}]{Riechers2008}
{Riechers}, D.~A., {Walter}, F., {Brewer}, B.~J., {Carilli}, C.~L., {Lewis},
  G.~F., {Bertoldi}, F., \& {Cox}, P. 2008, \apj, 686, 851

\bibitem[{{Riechers} {et~al.}(2006){Riechers}, {Walter}, {Carilli}, {Knudsen},
  {Lo}, {Benford}, {Staguhn}, {Hunter}, {Bertoldi}, {Henkel}, {Menten},
  {Weiss}, {Yun}, \& {Scoville}}]{Riechers2006}
{Riechers}, D.~A., {et~al.} 2006, \apj, 650, 604

\bibitem[{{Riechers} {et~al.}(2009){Riechers}, {Walter}, {Bertoldi}, {Carilli},
  {Aravena}, {Neri}, {Cox}, {Wei{\ss}}, \& {Menten}}]{Riechers2009}
---. 2009, \apj, 703, 1338

\bibitem[{{Solomon} {et~al.}(1997){Solomon}, {Downes}, {Radford}, \&
  {Barrett}}]{Solomon1997}
{Solomon}, P.~M., {Downes}, D., {Radford}, S.~J.~E., \& {Barrett}, J.~W. 1997,
  \apj, 478, 144

\bibitem[{{Solomon} \& {Vanden Bout}(2005)}]{Solomon2005}
{Solomon}, P.~M., \& {Vanden Bout}, P.~A. 2005, \araa, 43, 677

\bibitem[{{Steidel} {et~al.}(1999){Steidel}, {Adelberger}, {Giavalisco},
  {Dickinson}, \& {Pettini}}]{Steidel1999}
{Steidel}, C.~C., {Adelberger}, K.~L., {Giavalisco}, M., {Dickinson}, M., \&
  {Pettini}, M. 1999, \apj, 519, 1

\bibitem[{{Tacconi} {et~al.}(2006){Tacconi}, {Neri}, {Chapman}, {Genzel},
  {Smail}, {Ivison}, {Bertoldi}, {Blain}, {Cox}, {Greve}, \&
  {Omont}}]{Tacconi2006}
{Tacconi}, L.~J., {et~al.} 2006, \apj, 640, 228

\bibitem[{{Tacconi} {et~al.}(2008){Tacconi}, {Genzel}, {Smail}, {Neri},
  {Chapman}, {Ivison}, {Blain}, {Cox}, {Omont}, {Bertoldi}, {Greve},
  {F{\"o}rster Schreiber}, {Genel}, {Lutz}, {Swinbank}, {Shapley}, {Erb},
  {Cimatti}, {Daddi}, \& {Baker}}]{Tacconi2008}
---. 2008, \apj, 680, 246

\bibitem[{{Tacconi} {et~al.}(2010){Tacconi}, {Genzel}, {Neri}, {Cox}, {Cooper},
  {Shapiro}, {Bolatto}, {Bouch{\'e}}, {Bournaud}, {Burkert}, {Combes},
  {Comerford}, {Davis}, {Schreiber}, {Garcia-Burillo}, {Gracia-Carpio}, {Lutz},
  {Naab}, {Omont}, {Shapley}, {Sternberg}, \& {Weiner}}]{Tacconi2010}
---. 2010, \nat, 463, 781

\bibitem[{{Walter} {et~al.}(2004){Walter}, {Carilli}, {Bertoldi}, {Menten},
  {Cox}, {Lo}, {Fan}, \& {Strauss}}]{Walter2004}
{Walter}, F., {Carilli}, C., {Bertoldi}, F., {Menten}, K., {Cox}, P., {Lo},
  K.~Y., {Fan}, X., \& {Strauss}, M.~A. 2004, \apjl, 615, L17

\bibitem[{{Walter} {et~al.}(2009){Walter}, {Riechers}, {Cox}, {Neri},
  {Carilli}, {Bertoldi}, {Weiss}, \& {Maiolino}}]{Walter2009}
{Walter}, F., {Riechers}, D., {Cox}, P., {Neri}, R., {Carilli}, C., {Bertoldi},
  F., {Weiss}, A., \& {Maiolino}, R. 2009, \nat, 457, 699

\bibitem[{{Wei{\ss}} {et~al.}(2007){Wei{\ss}}, {Downes}, {Neri}, {Walter},
  {Henkel}, {Wilner}, {Wagg}, \& {Wiklind}}]{Weiss2007}
{Wei{\ss}}, A., {Downes}, D., {Neri}, R., {Walter}, F., {Henkel}, C., {Wilner},
  D.~J., {Wagg}, J., \& {Wiklind}, T. 2007, \aap, 467, 955

\bibitem[{{Wei{\ss}} {et~al.}(2005){Wei{\ss}}, {Downes}, {Walter}, \&
  {Henkel}}]{Weiss2005b}
{Wei{\ss}}, A., {Downes}, D., {Walter}, F., \& {Henkel}, C. 2005, \aap, 440,
  L45

\bibitem[{{Yao} {et~al.}(2003){Yao}, {Seaquist}, {Kuno}, \& {Dunne}}]{Yao2003}
{Yao}, L., {Seaquist}, E.~R., {Kuno}, N., \& {Dunne}, L. 2003, \apj, 588, 771

\bibitem[{{Zhu} {et~al.}(2003){Zhu}, {Seaquist}, \& {Kuno}}]{Zhu2003}
{Zhu}, M., {Seaquist}, E.~R., \& {Kuno}, N. 2003, \apj, 588, 243

\end{thebibliography}

%

\clearpage

\end{document}